# Status of the 0.8 Teraflops Supercomputer at Columbia.[*]

I. Arsenin [a], D. Chen[a], N. Christ[a], R. Edwards[d], A. Gara[b], S. Hanson[c], C. Jung[a], A. Kahler[a], A.D. Kennedy[d], G. Kilcup[e], Y. Luo[a], C. Malureanu[a], R. Mawhinney[a], J. Parsons[b], J. Sexton[f], C. Sui[a], and P. Vranas[a][†]

[a]Dept. of Physics, Columbia University, New York, NY 10027

[b]Nevis Laboratory, Columbia University, Irvington on Hudson, NY 10533

[c]Fermilab, P.O. Box 500, Batavia, IL 60510

[d]SCRI, Florida State University, Tallahassee, FL 32306-4052

[e]Dept. of Physics, Ohio State University, 174 West 18th Ave, Columbus, Ohio 43210

[f]Dept. of Mathematics, Trinity College, Dublin 2, Republic of Ireland

The first stage in the construction of the 0.8 Teraflops Supercomputer at Columbia, a working, two node parallel computer, has been successfully completed. The next stage, a 512 node, 26 Gigaflops prototype, is in its final construction phase. A general description and current status of the hardware and software is presented.

## 1. Introduction

The 0.8 Teraflops Supercomputer [1] consists of 16,384 nodes connected in a $16^3 \times 4$ 4-D mesh. Each node delivers a peak speed of 50 Mflops, has 0.5 Mwords of DRAM, and 25/50 MHz bit–serial communications. The machine is single precision, except for global sums that can be of arbitrary precision. The cost is 3 million U.S. dollars.

Each node is mounted on a daughterboard and consists of a Texas Instruments Digital Signal Processor (DSP) chip, DRAM chips, and a custom made chip called the Node Gate Array (NGA) that provides an efficient memory and communication interface. Sixty–four nodes are housed on a motherboard. Eight motherboards are plugged into a backplane and are housed in a crate that provides power and cooling. Thirty–two crates are cabled together to make up the machine (8 racks of 4 vertically stacked crates).

## 2. Programming Environment

A well supported programming environment will be available to the user. The time non-critical portions of the code can all be written in C, using the C compiler that accompanies the Texas Instrument DSP and a few C-level routines that will perform the communications and, if needed, improve the memory access timings. As with most machines, the time critical portions of the code must be "hand-coded" for optimal performance. This can be done by using the standard Texas Instruments assembler and the memory mapped NGA. The Wilson and staggered inverters have already been "hand-coded" and they sustain more than 30% of the peak speed on $2^4$ lattices per node.

Code development tools include the Texas Instruments DSP debugging tools that can be used to debug the serial portion of the code. A debugging kernel that can be used to debug the parallel part of the code is being planned. Also, few–node machines dedicated to code development will be built.

Finally, the lattice sizes directly available to the user are multiples of $16^3 \times 4$. Other sizes can be achieved at the software level using the pass–through capability of the NGA or at the hardware level by recabling the machine.

## 3. Hardware

**1) NGA:** The most important component of the machine is the Node Gate Array (NGA). This is the device that links all other components together in an efficient way. It provides interfaces to the DSP, DRAM, and to each of the eight nearest neighbors via 8, bit–serial lines. It is an Application Specific Integrated Circuit (ASIC), about $1.25 \times 1.25$ inches in size and has 208 pins. It consists of $260,000$ transistors (as many as in an 80286 chip). For more details regarding the design of the NGA the reader is referred to [1].

---
[*]Research was supported by the U.S. Dept. of Energy.
[†]Speaker

The NGA was entirely designed and tested on a PC and Sparc-10 workstation at Columbia, using the Viewlogic software tools. Software models of the DSP and DRAM, available from Logic Modeling Co., were "connected" with the software model of the NGA to provide a fully operational model of a single node. In this model each serial line along one direction was connected with the one in the opposite direction.

In order to test the NGA, a quarter of a million cycles of code including conjugate gradient inverters for staggered and Wilson fermions were run. This code was run on the single node model using the Viewlogic simulator to first test the logic of the design. Next, the design was compiled to an actual circuit of gates. The test code was run again using the simulator but this time to test the hardware operation of this circuit. At this stage the software simulates the operation of the actual chip as if it was already manufactured. In particular, for a given set of external parameters, such as temperature, voltage and manufacturing quality, the simulator reproduces the delays the signals encounter as they propagate through the circuit. The chip must work over a range of these parameters since in reality they may vary. Collectively this range is parametrized by a quantity called derating. The chip was tested for three deratings covering the expected range of parameters.

To understand one of the main sources of malfunctions that occurred as the derating was varied consider a signal propagating in the circuit. At the rising edge (re) of a clock cycle, this signal is expected to have arrived at a given point in the circuit where an "edge sensitive" device such as a flip-flop is located. The flip-flop sets its output (Q) for the whole cycle to have the same value as the value its input had at the beginning of the cycle. A situation where a signal is on time, or is "missed" because it arrived too early or too late is shown in fig. 1. Since a signal is a combination of many other signals, the critical path that caused the signal to arrive too late or too early had to be found in order to identify and fix the cause of the problem. Because of the complexity of the circuit and the fact that the simulator needed about a second to simulate just a few chip cycles this was a non–trivial and time consuming task.

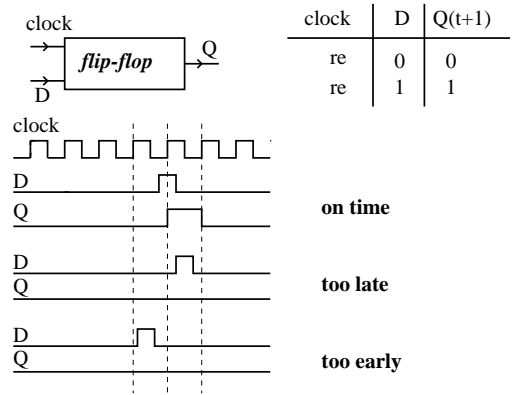

**Figure 1.** Signal arriving at a flip-flop

In order for the chip to be manufactured, the final design must be accompanied by a file containing test vectors. A test vector is a collection of input signals and the corresponding outputs they produce. Test vectors are used by the manufacturer to test the proper operation of the device as it is transcribed from a software model to an actual chip. The test vectors were produced by running the quarter million cycles worth of code using the simulator. However, these test vectors were not in accordance with the manufacturer's requirement that in every 20 ns cycle there must be a 5 ns window (called strobe window) where no output signal is changing.

This problem was fixed by using the fact that a set of output signals that go to a peripheral device is important only when a corresponding output control signal indicates that the peripheral device should "pay attention". By design, these output signals do not change for a few ns before this (setup time). The NGA has three types of control signals, one for the DSP, one for the DRAM and one for the serial communication signals. For each control signal a strobe window can be created by keeping the corresponding output signals only during the setup time. The rest of the time these signals are not important and can therefore be neglected. Still, this is not enough because the three control signals occur at different points in the cycle and therefore their corresponding strobe windows do not overlap. To avoid this final problem the test vectors were split into three groups — one for each set of signals. Special software was developed to analyze and manipulate the test vectors produced by the simulator and bring them to their final form of three groups, each with its own strobe window.

The design and testing of the NGA have been completed and several chips have been manufactured and delivered by the Atmel Corporation.

**2) Daughterboard:** The NGA, the DSP, and the DRAM are mounted on the daughterboard. The daughterboard is a $2.6 \times 1.75$ inch, 6–layer custom made printed circuit board. Several have been manufactured and delivered.

**3) Two node prototype:** A prototype motherboard that houses two daughterboards was designed as an intermediate testing step. It is a 4–layer custom made printed circuit board. Several have been manufactured and delivered. Using this, two daughterboards and some basic software (to be discussed later), a two node prototype parallel computer was built and was operational at the beginning of August. The careful design and exhaustive testing of the various components resulted in a prototype that was basically free of errors and worked (running the staggered conjugate gradient algorithm) within one week from the day its assembly began.

At present seven nodes have been assembled and have passed an exhaustive set of tests. In particular, the code that produced the test vectors was run in a tight loop for several hours with no errors. All tests were run with the high, 1.2 Mwords/sec, serial communication speed, at room temperature and with no air blowing at the node. The peak power consumption of a single node was measured to be less than 3.5 Watts. Furthermore, all nodes also ran successfully with a 64 MHz clock (28% faster than the design 50 MHz peak speed).

**4) Motherboard:** The motherboard is a $14.5 \times 20.5$ inch, 8–layer custom made printed circuit board that houses 64 nodes. One node called node 0, is mounted directly on the motherboard. The remaining 63 are mounted on daughterboards and are plugged into the motherboard using SIMM–sockets. Node 0 is special in that it has more memory (4 Mwords), is connected to a PROM from which it will boot, and controls two SCSI chips and a Programmable Array Logic (PAL) switch that connect it with the control/diagnostic network (to be described later). The motherboard schematic is complete and it is currently being laid-out.

**5) Backplane:** The backplane is a $14.5 \times 24$ inch, 12–layer custom made printed circuit board with sockets to plug in eight motherboards. It provides each motherboard with power and connections to all necessary signals. The design has been completed and is currently being manufactured.

**6) Crate:** The crate is a cabinet that contains the 8 motherboards, the backplane and power supplies at the rear, and muffin fans and a cold water radiator, when mounted in a rack, at the bottom. It is currently being manufactured by our supplier.

## 4. Networks

There are two independent networks:

**1) The physics network** connects all nodes in a $16^3 \times 4$ hypercubic mesh using bit–serial lines. It is software programmable for a 25/50 MHz (0.6/1.2 Mwords/sec) transmission rate. This network will be used by the physics code for nearest neighbor communication. It will also be used to "unload" lattices from the whole machine to the node 0 of each motherboard and from there via a SCSI link to disk. The proper operation of this network was tested using the prototype two node computer where the signal along one of the serial directions was carried by a 10 foot long cable. It worked successfully at 50 MHz. The largest sized cable in the full machine will be about 6 feet.

**2) The control/diagnostic network** is completely independent of the physics network and does not involve the NGA or DRAM. Because of this it is less sensitive to hardware failures. It will provide the front–end connection to the host workstation (SUN Sparc–10), diagnostic information, and will be used to load and initiate execution of programs as well as unload small amounts of data. It consists of two parts. The first part is a SCSI tree with the host-workstation at the root, connecting the 256 motherboards and a number of disks. A motherboard is connected to the SCSI tree via two SCSI chips controlled by node 0. The second part is a tree on each motherboard that connects node 0 with the remaining 63 nodes via their DSP serial ports. Node 0 can select any, or all, of the 63 nodes via a programmable PAL switch. A SCSI link has a bandwidth of 2.5 Mwords/sec and the DSP serial port has a bandwidth of 0.3 Mwords/sec.

## 5. Software

When the power is turned on, all the DSP's start running a boot–loader code that has been imprinted on the chip by the manufacturer. All nodes 0 have a pin set so that the boot–loader will read boot–code from the PROM into memory and start executing it. This code sets up the DSP serial port and the PAL switch and broadcasts a copy of itself to the serial ports of the other 63 nodes. These nodes have a pin set so that the boot-loader waits until code arrives at their serial ports. Once the copy of the boot–code sent by node 0 arrives, the boot-loader puts it into memory and starts executing it. The boot–code performs some diagnostic tests and then branches to a kernel.

The kernel on nodes 1–63 services a set of fundamental requests that arrive at the DSP serial port, such as write, read and execute. The kernel on node 0 is more complex since it has to control the two SCSI chips, the PAL switch and handle the routing and service requests along the SCSI tree, as well as along the local DSP serial port tree. The kernel expects to receive requests according to a predefined communications protocol. This consists of packets with a maximum size of 512 words (so that they can be easily buffered in the DSP chip memory which is 2 Kwords long) that contain a "payload" of data, as well as routing and control information. Once the code has branched to the kernel, the machine is accessible from the host workstation. The host communicates with the machine through a front–end program that exchanges packets via the SCSI connection.

The communications protocol has been finalized. The kernel for nodes 1–63 has been written and was used to boot a single node and download and run some of the code that tested the NGA, as well as to boot the two node prototype. The SCSI driver for both the SUN host workstation and the motherboard SCSI chips, the fundamental part of the kernel for node 0, and a program loader for the SUN have also been written. They have been tested successfully on existing commercial hardware and are ready to be ported to the actual hardware when it is assembled. The remainder of the kernel for node 0 has been designed and is currently being written. Also, physics code is currently under development.

## 6. Timetable

We expect to have the 26–Gigaflops 512–node prototype working by the end of 1995 and hope to have the full machine finished a year later.

## 7. Acknowledgments

We would like to thank the the Columbia summer students X. Chen, S. Kasow and C. Lazaroiu for their help.